\documentclass[utf8x]{article-hermes}

\newcommand{\tool}[1]{{\texttt{#1}}}
\usepackage{multirow}
\usepackage{covington}
\usepackage{hyperref}
\usepackage{xcolor}
\usepackage{tabularx}

\submitted[01/09/2015]{TAL~55-03}{3}

\title[Analyse lexicale de paroles schizophrènes]{Analyse lexicale outillée de la parole transcrite de patients schizophrènes}

\author{Maxime Amblard\fup{*} \andauthor Karën Fort\fup{**}  \\
Caroline Demily\fup{***} \andauthor Nicolas Franck\fup{***} \andauthor Michel Musiol\fup{****}}

\address{
\fup{*} LORIA, UMR 7503, Université de Lorraine, CNRS, Inria\\
\fup{**} Université Paris-Sorbonne / EA STIH\\
\fup{***} UMR 5229 CH le Vinatier CNRS et Université Lyon 1\\
\fup{****}ATILF, UMR 7118, Université de Lorraine, CNRS
}

\resume{
Cet article détaille les résultats d'analyses réalisées sur la transcription d'entretiens avec des patients schizophrènes, aux niveaux de la production orale (disfluences) et du lexique (morpho-syntaxe et lemmes). L'étude s'inscrit dans le cadre d'un projet plus large qui prévoit d'autres niveaux d'analyse (syntaxique et du discours), les résultats obtenus devant nous permettre de réfuter ou d'identifier de nouveaux indices linguistiques présents dans la manifestation d'un dysfonctionnement à ces différents niveaux.
Le corpus traité contient plus de 375~000 mots, son analyse a donc nécessité l'utilisation d'outils de traitement automatique des langues (TAL) et de textométrie. Nous avons en particulier séparé le traitement des disfluences du traitement lexical, ce qui nous a permis de montrer que si les schizophrènes produisent davantage d'achoppements et de répétitions (disfluences) que les témoins, la richesse de leur lexique n'est pas significativement différente.
}

\abstract{
This article details the results of analyses we conducted on the discourse of schizophrenic patients, at the oral production (disfluences) and lexical (part-of-speech and lemmas) levels. This study is part of a larger project, which includes other levels of analyses (syntax and discourse). The obtained results should help us rebut or identify new linguistic evidence participating in the manifestation of a dysfunction at these different levels. The corpus contains more than 375,000 words, its analysis therefore required that we use Natural Language Processing (NLP) and lexicometric tools. In particular, we processed disfluencies and parts-of-speech separately,  which allowed us to demonstrate that if schizophrenic patients do produce more disfluencies than control, their lexical richness is not significatively different.
}

\motscles{
schizophrénie, disfluences, morpho-syntaxe, lemmes, discours pathologique}


\begin{document}

\maketitlepage

\section{Introduction}
De nombreuses études ont porté sur la définition, l'implémentation et l'évaluation d'outils pour analyser les pratiques langagières. Leurs motivations s'inscrivent dans des tâches bien définies, mais la question de la validité cognitive des théories et modèles est souvent reléguée à des ouvertures plus qu'à de véritables arguments.
S'il apparaît évident qu'il s'agit d'une question complexe, il n'en est pas moins nécessaire d'interroger ces propositions sous l'angle du fonctionnement cognitif.

Une manière d'appréhender cette problématique est de s'intéresser à des manifestations explicites du fonctionnement cognitif, par exemple 
des dysfonctionnements plutôt qu'à des usages supposés normaux. Ici, nous nous intéressons à l'étude de la réalisation de phénomènes spécifiques chez les schizophrènes au travers de leur pratique langagière.
 Ces phénomènes sont analysés comme des dysfonctionnements dans la planification du discours, symptôme d'un dysfonctionnement cognitif~\cite{rebuschihal-00910725}.
  Nous ne tentons pas de définir comment le cerveau produit ce dysfonctionnement mais comment et où ce dysfonctionnement apparaît du point de vue linguistique. Une fois ce dysfonctionnement circonscrit, il deviendra possible de travailler à un modèle pour en rendre compte précisément.
Dans un mouvement inverse, étudier ces dysfonctionnements apparaissant dans la langue permet de donner une validité cognitive aux outils capables de les identifier.

Ces travaux s'inscrivent dans le cadre d'une étude large portant sur les pratiques langagières de patients schizophrènes.
Le matériel de cette étude provient principalement d'entretiens semi-dirigés par des psychologues. Ces entretiens sont définis pour minimiser l'apport du psychologue, laissant une place importante à la parole du patient. Il s'agit de recueillir l'expression de sa pensée, pour procéder à une analyse, en général psychologique ou psychiatrique. Mais ce n'est pas notre sujet. Nous appréhendons ce matériau comme l'expression d'une pensée en action, au sens cognitif, et l'analysons du point de vue de la pratique langagière.

Dans la continuité des travaux de~\cite{Chaika74} et~\cite{Fromkin1975498} qui, les premiers, ont cherché à mettre en avant des indices spécifiques à la capacité langagière des schizophrènes. Ils se sont appuyés sur l'hypothèse forte que la forme de l'expression de leur pensée véhiculait des informations sur les processus cognitifs en \oe uvre. En même temps, les schizophrènes manifestent des particularités.
Si Chaïka s'intéresse à la capacité d'appliquer des règles syntaxiques,
\cite{Landre92} rapporte que les schizophrènes font le même type d'erreurs que les aphasiques, ce qui les conduit à donner une origine extralangagière au dysfonctionnement, le positionnant à un plus haut niveau cognitif.
\cite{Besche96} ont étudié la pratique lexicale des patients schizophrènes pour également réfuter l'idée qu'ils auraient un trouble généralisé de traitement du contexte, à nouveau inscrivant les dysfonctionnements à un niveau cognitif plus élevé.

Cependant, ces études restent très limitées, tant dans le nombre de patients pris en considération que dans l'ampleur des phénomènes analysés. En général, et au vue de la difficulté de rencontrer de tels patients, ces études incluent seulement une vingtaine de participants. Par ailleurs, les moyens tant matériels que théoriques à la disposition des auteurs les contraignent à réaliser à la main des tests relativement peu avancés. Nous ne souhaitons aucunement remettre en cause leurs méthodes, mais utiliser des outils et méthodologies développés dans le cadre du traitement automatique des langues (TAL) sur ces données particulières.

Dans une première partie, nous revenons sur le contexte de cette étude, tant du point de vue de son organisation que de son contexte scientifique. Nous en précisons également le cadre et les limites. Puis nous présentons le corpus en revenant sur sa constitution et les difficultés de la création d'une telle ressource. Enfin, nous détaillons les outils utilisés et les résultats obtenus sur le corpus en analysant les achoppements et les répétitions (disfluences), les catégories morpho-syntaxiques, et les lemmes produits. Nous proposerons ensuite une brève analyse textométrique avant de conclure.


\section{Contexte de l'étude}

Si nous disposons aujourd'hui de nombreuses références d'articles traitant du sujet de la production langagière des schizophrènes, il n'est pas aussi simple d'en tirer des conclusions. Outre que ces articles proviennent de domaines variés (psychologie, médecine, linguistique, etc.) et qu'ils sont plus ou moins récents et plus ou moins facilement disponibles selon les traditions de chaque domaine, les conditions des expériences décrites sont d'une telle variabilité qu'il est difficile d'en mettre les résultats en cohérence. En effet, les tailles de corpus et les protocoles varient énormément, la langue diffère, les patients sont pour certains en remédiation (et sous-traitement), d'autres non. Enfin, les résultats sont comparés dans certains cas à des témoins et dans d'autres à des patients souffrant d'autres désordres ou pathologies.

La méta-étude de Brendan Maher~\cite{Maher1972} est très intéressante de ce point de vue, car l'auteur signale les biais de telle ou telle étude ou leurs différences. S'il présente ensemble les résultats concernant les répétitions et ceux concernant la richesse lexicale, il est l'un des rares à les distinguer. Ses conclusions sur les répétitions, déduites des TTR (\textit{Type-Token Ratio}), sont relativement claires : les patients schizophrènes ont un TTR inférieur, ce qui signifierait qu'ils se répètent davantage. Des perturbations du discours des schizophrènes (achoppements, répétitions) ont également été observées par d'autres, notamment dans~\cite{Feldstein1962} et~\cite{Kremen2003}.

Par ailleurs,~\cite{Maher1972} cite, tout en  émettant  certaines réserves (les données étant trop limitées), des résultats qui montreraient que les patients schizophrènes utiliseraient un vocabulaire plus restreint.
  Une étude portant sur les familles de schizophrènes et détaillée dans~\cite{DeLisi2001} montre elle-aussi que les schizophrènes chroniques utilisent significativement moins de mots que les témoins. 

  L'analyse que nous présentons vise à vérifier ces résultats, sur une cohorte relativement large.
  Elle est à notre connaissance la seule portant sur des patients francophones, surtout, elle a été réalisée à l'aide d'outils de TAL au niveau état de l'art.
  %
 Il est à noter que, comme les travaux que nous présentons s'inscrivent dans un projet plus général,
 le corpus sur lequel nous travaillons est partagé avec d'autres recherches. Le protocole utilisé couvre lui l'ensemble du projet.
 En particulier, nous mesurons les capacités neuro-cognitives par une série de tests avant l'entretien et au cours de certains, nous enregistrons le comportement oculomoteur du patient avec un oculomètre (\textit{eye-tracker}) et/ou l'activité encéphale par  électro-encéphalographe (EEG). Dans cet article, nous n'utilisons que les enregistrements sonores transcrits des entretiens.

\section{Constitution du corpus}
\label{corpus}
Notre étude, comme la plupart de celles sur les pratiques langagières des patients schizophrènes, est confrontée à de nombreux obstacles pour la constitution du corpus.

\subsection{Répartition des sujets}

Aux vues des difficultés pour identifier les patients et les faire intervenir dans l'étude, notre corpus a été constitué en plusieurs phases, dans différents centres hospitaliers. Dans l'analyse présentée ici, nous considérons les résultats de deux cohortes comportant en tout 80 sujets qui se répartissent en 49 schizophrènes et 31 témoins. Le tableau~\ref{tablerepartitionsujets} présente la ventilation des sujets en fonction de leur type (schizophrène ou témoin) et de leur sexe.

Le corpus est divisé en deux cohortes, correspondant aux villes des unités médicales spécialisées des recueils. Par respect pour la confiance accordée par les patients, nous anonymisons ces noms de villes en Ville1 et Ville2. Le recueil de Ville1 a été réalisé par une psychologue pour les patients et trois psychologues pour les témoins, 
et celui de Ville2 
par les deux mêmes psychologues pour les patients et les témoins.

Le sous-corpus Ville1 a été constitué au second semestre 2013. Il est composé de 18 patients diagnostiqués schizophrènes, en remédiation et sous traitement, ainsi que de 23 témoins.
Le sous-corpus Ville2 a été constitué au printemps 2002. Il est composé de 31 patients diagnostiqués schizophrènes en remédiation et sous traitement, à l'exception de sept d'entre eux (qui n'étaient pas sous traitement), et de 8 témoins.
%

\begin{table}[h]
\begin{center}
\begin{tabular}{|l|cc|c|cc|c|l|}
\hline
& \multicolumn{3}{l|}{corpus Ville1}& \multicolumn{3}{l|}{corpus Ville2}& total\\
\hline
& hommes & femmes & total & hommes & femmes &total &\\
\hline
schizophrènes & &   & &  &  &  &\\
~~sous traitement &15 & 3 &18 & 21 & 3 & 24 &\\
~~sans traitement & 0 &0 &0 & 1 & 6 & 7 &\\
total & &   &18 &  &  & 31 &49\\

\hline
témoins &15&8 &23&4 & 4& 8 &31\\
\hline
total & 30 & 11 &41&26 &13&39&80\\
\hline
\end{tabular}
\end{center}
\caption{Répartition des sujets dans le corpus en fonction des cohortes et du sexe.\label{tablerepartitionsujets}}
\end{table}


\subsection{Protocole de collecte}


L'interaction choisie pour cette étude s'organise autour d'un entretien semi-dirigé conduit par un psychologue. Ce type d'entretien est bien défini dans la communauté psychologique et psychanalytique (bien que la terminologie puisse varier) : il s'agit pour le psychologue de maintenir une interaction dans laquelle l'interlocuteur parle librement de lui-même. Pour cela il revient sur son environnement matériel direct, ses relations humaines dans son cadre, ainsi qu'à l'extérieur de son cadre. Le psychologue n'est en aucun cas personnellement engagé dans l'interaction, et sa contribution principale est de relancer l'échange ou de préciser certains éléments. 

Par ailleurs, lors de la constitution du sous-corpus Ville1, les sujets ont passé une série de tests permettant de mesurer certaines compétences cognitives. Les tests choisis sont classiques, au sens où ils sont régulièrement utilisés dans la littérature pour des analyses similaires.
Les trois tests psychocognitifs choisis mesurent les capacités de mémoire à court terme, d'attention, et la mémoire de travail :
\begin{enumerate}
\item le \textit{Wechsler Adult Intelligence Scale-III} (mesure du quotient intellectuel, ou QI),
\item le \textit{California Verbal Learning Test} (capacité cognitive et de stratégie),
\item  le \textit{Trail Making Test} (dépréciation de la flexibilité cognitive et de l'inhibition, déficit qui peut affecter la vitesse du système perceptif-moteur, la flexibilité spontanée ou la flexibilité de réaction).
\end{enumerate}

Dans le présent article, nous n'utiliserons que les résultats du test de QI.
%
%
Il nous semble important d'insister sur le fait que le protocole stipule explicitement que le contenu de l'entretien ne peut et ne doit pas être utilisé ni pour, ni contre le patient.
Le fait de ne pas utiliser le contenu contre le patient leur permet une certaine liberté d'expression, et dans un mouvement inverse ne pas l'utiliser pour eux limite la tentation de renvoyer une image trop positive d'eux-mêmes dans le contexte hospitalier.

\subsection{Transcription de la parole}

Nous récupérons les enregistrements des entretiens sous forme de fichier sonore mp3.
Ils sont alors transcrits. Nous considérons la transcription comme le premier niveau d'annotation de la ressource. Les deux sous-corpus ayant été constitués à des moments très écartés dans le temps (plus de 10 ans les séparent), les processus de transcription n'ont pas pu être les mêmes. Cependant, dans les deux cas, les transcriptions ont été réalisées par plusieurs annotateurs. Il s'est agit du ou de la psychologue qui a mené tout ou partie des entretiens, ainsi que d'une seconde personne. L'investissement en temps sur cette tâche étant limité, les transcriptions n'ont malheureusement pas pu être réalisées en parallèle.

Il est important de noter que les transcripteurs n'ayant pas connaissance de l'utilisation de leur travail pour des tâches de TAL, n'ont probablement pas pu influencer les résultats dans un sens ou un autre. Les annotateurs ont suivi les recommandations de base fournies avec \tool{Transcriber} pour une transcription fine et la transcription a été post-traitée suivant les préconisations de~\cite{BlancheBenveniste1987}.
Nous avons réalisé une relecture partielle \textit{a posteriori} pour identifier les unifications d'annotations minimales à apporter à l'ensemble de la ressource par une série de scripts de normalisation tant sur le codage du texte, le format des fichiers que les annotations elles-mêmes.

En moyenne, les entretiens du sous-corpus de Ville1 sont constitués de 552,73 tours de parole, alors que les entretiens du sous-corpus Ville2 en contiennent 234,5. L'ensemble du corpus comprend 31~575 tours de parole, soit environ 375~000 mots. Le tableau~\ref{tablecorpusTdPmot} présente la répartition en tours de parole et en mots de l'ensemble du corpus. Il faut noter que, du point de vue du TAL, ce corpus reste de taille modeste. Cependant, nous considérons qu'il atteint une taille raisonnable pour l'utiliser, au vue de sa spécificité, aspect sur lequel nous revenons dans la section suivante.

\begin{figure}[ht]
\begin{center}
\includegraphics[scale=0.4]{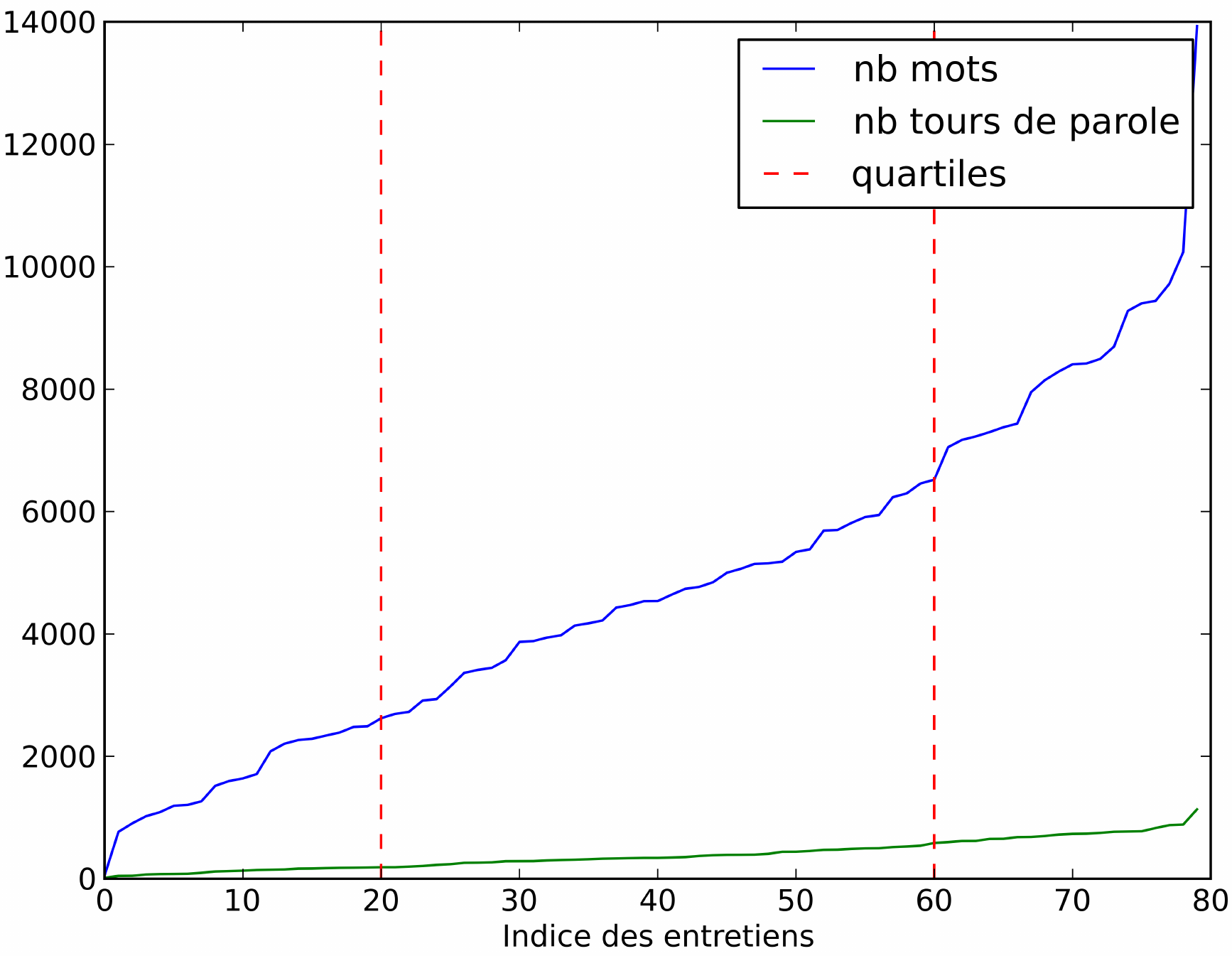} \\
\end{center}
\caption{Distribution des tailles des entretiens en nombre de mots (en bleu) et en nombre de tours de parole (en vert), ainsi que les quartiles (en rouge).}
\label{fig:distriParTaille}
\end{figure}

La figure~\ref{fig:distriParTaille} illustre la distribution des tailles des entretiens en nombre de mots et de tours de parole, pour les deux cohortes.
Les quartiles apparaissent en rouge. On constate que les entretiens contiennent pour la moitié d'entre eux entre 2~500 et 6~000 mots. Par ailleurs, le nombre de tours de parole est relativement constant dans le corpus. On calcule qu'en moyenne un entretien est composé de 393 tours de parole pour 4~792 mots.
Le caractère spécifique de l'entretien semi-directif apparaît dans le corpus~: le psychologue produit le même nombre de tours de parole que le sujet, pour un volume de mots très inférieur.
Par exemple, dans le sous-corpus Ville1, le nombre de tours de parole des schizophrènes et des psychologues devant un schizophrène est le même, alors que le volume de mots des schizophrènes est 1,54 fois plus important que celui des psychologues. Les témoins du sous-corpus de Ville1 ne présentent pas cette caractéristique, mais une analyse plus fine des entretiens montre que pour six d'entre eux, les témoins ont été réticents à prendre la parole.

\begin{table}[h]
\begin{center}
$
\begin{array}{|l|ll|ll|ll|ll|}
\hline
&\multicolumn{4}{c|}{$corpus Ville1$} & \multicolumn{4}{c|}{$corpus Ville2$} \\
\hline
&\multicolumn{2}{c|}{$nb tours$} &\multicolumn{2}{c|}{$nb mots$}&\multicolumn{2}{c|}{$nb tours$} &\multicolumn{2}{c|}{$nb mots$}\\
\hline
S &3~863& \multirow{2}{*}{ 11~145} &46~859 &\multirow{2}{*}{119~762} & 4~062 & \multirow{2}{*}{ 4~433 } & 66~725  &\multirow{2}{*}{ 79~081}\\
T& 7~282 && 72~903 & &371 & &12~356&\\
\hline
P + S& 3~819&\multirow{2}{*}{11~517} & 30~293 & \multirow{2}{*}{138~571} & 4~098 & \multirow{2}{*}{4~480} & 33~686 & \multirow{2}{*}{37~842}\\
P + T& 7~698&& 108~278 & & 382& & 4~156 &\\
\hline
total & \multicolumn{2}{r|}{22~662}&\multicolumn{2}{r|}{258~333} & \multicolumn{2}{r|}{8~913} & \multicolumn{2}{r|}{116~923}\\
\hline
\end{array}
$
\end{center}
\caption{Décomposition du corpus en sous-corpus, en nombre de tours de parole et nombre de mots, en fonction du type d'interlocuteur:\label{tablecorpusTdPmot} S (schizophrènes), T (témoins), P + S (psychologue avec un schizophrène), P + T (psychologue avec un témoin).}
\end{table}

Notre étude se focalise sur des aspects relevant du lexique et de la morpho-syntaxe. Aussi nous n'avons pas exploité les aspects 
phonétiques, comme le temps de parole ou la vitesse d'élocution des locuteurs, ni le recouvrement des tours de parole.
Ces données restent cependant disponibles dans le corpus pour une étude ultérieure.

\subsection{Difficultés d'accès aux patients}

Le nombre de 80 sujets peut sembler limité, mais la constitution d'une telle ressource implique de surmonter de nombreuses difficultés, en particulier pour accéder aux patients. De ce fait, disposer d'une cinquantaine de transcriptions d'entretiens avec des schizophrènes représente déjà un corpus significatif.

Pour s'entretenir avec une personne prise en charge en milieu hospitalier, il est nécessaire d'obtenir une autorisation du CPP (Comité de Protection de la Personne) de la région de l'établissement. Les demandes déposées contiennent explicitement et exactement le protocole. L'instruction du dossier requiert plusieurs mois et demande la contraction d'une assurance (pour prendre en charge les possibles dommages). Ces assurances augmentent considérablement les budgets nécessaires à ce type d'expérience. Une fois les accords obtenus, il n'est alors plus possible de modifier les protocoles.

Mais ce qui rend complexe la constitution d'une telle ressource est principalement la difficulté de faire participer les patients. Plusieurs problèmes se posent. Il faut d'abord identifier, au sein d'un service, les patients répondant aux critères de l'étude et en capacité d'interagir avec une personne tierce au service. Puis il faut, au sein de cette population,  trouver les patients qui acceptent de participer. Une première réticence vient du fait qu'il n'y a pas de conséquence positive, en terme médical, à participer à l'étude. Il faut ajouter à cela des inquiétudes compréhensibles des patients concernant la possible publication de leur histoire, bien qu'une anonymisation soit garantie.

Par ailleurs, le protocole requérant de passer des tests psycho-cognitifs et un entretien, le temps nécessaire est de l'ordre de deux heures, ce qui est relativement élevé. Ce n'est pas tant la disponibilité des patients qui est alors en jeu, que leur aptitude à rester concentrés. Lorsque le patient présente soudainement des difficultés, il faut convenir d'un second rendez-vous pour finaliser le protocole. La multiplication des rendez-vous génère des défections.
À titre d'exemple, lors de la phase de collecte des entretiens du sous-corpus Ville1 qui s'est déroulée dans d'excellente conditions matérielles et administratives, 45~\% (18) des patients contactés ont refusé de participer, 10~\% ont accepté un premier rendez-vous mais ne sont pas présentés au second, et 45~\% (18 sujets) ont participé à toute l'étude.


\subsection{Anonymisation}
\label{anonymat}
L'anonymisation d'un corpus est une tâche complexe qui recouvre plusieurs dimensions.
Nous avons dans un premier temps cherché à désidentifier le corpus~\cite{Meystre2010}, c'est-à-dire à identifier les entités nommées et à leur substituer des marqueurs neutres.

Afin de conserver la lisibilité des documents, nous avons cherché par lectures successives les catégories à masquer. Nous en avons identifié 10.
Concernant les personnes, nous identifions les noms et nous avons choisi de conserver une marque explicite du sexe pour le prénom, ce qui permet de lever des ambiguïtés dans plusieurs entretiens. De manière tout à fait classique~\cite{Grouin2013}, nous avons choisi d'utiliser une catégorie pour les institutions, qui ici sont nombreuses, car il est souvent fait mention d'hôpitaux ou de différents services d'un même hôpital. Cette analyse nous impose de conserver un identifiant unique pour chaque entité. Ainsi, si la première référence faite à l'hôpital X est substituée par \textit{institution3}, ce même hôpital restera \textit{institution3} dans tout l'entretien.
Il ne serait plus possible de comprendre les échanges sans cette contrainte. Cela étant, nous ne conservons cette cohérence de dénotation qu'à l'intérieur d'un même entretien et non sur l'ensemble du corpus. Par extension, cette propriété est conservée pour toutes les catégories.
Les entretiens étant situés dans une géographie particulière, nous utilisons également les catégories
\textit{pays}, \textit{département}, \textit{ville}, \textit{capitale}, \textit{montagne}. Ces catégories pourraient être amenées à évoluer, en particulier autours des relations ontologiques qu'elles entretiennent. Le choix a été fait de les fixer à cela pour cette phase du projet.
Enfin, nous avons ajouté une dernière catégorie rassemblant tous les autres cas : \textit{non\_pris\_en\_compte}.
Nous ne présentons pas ici de répartition de ces annotations d'anonymisation, mais elles apparaissent en très faible quantité.

Nous avons identifié un outil d'anonymisation performant, MEDINA~\cite{Grouin2013}, mais nous n'avons pu l'obtenir à temps pour des raisons administratives\footnote{Il faut, pour utiliser cet outil, faire signer sa licence par les laboratoires.}.
Nous avons donc implémenté une série de scripts en \tool{Python} basés sur les expressions régulières. Une intervention humaine reste nécessaire pour superviser l'application des scripts qui peuvent lever des exceptions en cas d'ambiguïté. Cette tâche étant réalisée une fois pour toute, nous avons choisi d'en privilégier la qualité, quitte à perdre en efficacité. À partir du résultat, nous avons donc procédé à une vérification par extraction automatique de toutes les positions potentiellement ambiguës. 

Nous nous sommes rendus à l'évidence, comme d'autres avant nous (notamment~\cite{Eshkol-Taravella2014}), que s'il nous était possible de cacher le prénom et le nom des personnes, les sujets exprimant de nombreuses informations tant sur leur histoire que sur leur famille, voire leur localisation, il nous est impossible de garantir un véritable anonymat.
Ainsi, dans l'un des entretiens, le patient explique qu'il a intégré une classe préparatoire dans une ville du Nord, avant de retourner s'inscrire à l'université dans une autre ville, suite à une dépression. Ces deux éléments peuvent paraître peu, mais mis ensemble, qui plus est en ajoutant d'autres informations sur sa famille disséminées dans l'entretien, il devient, sinon possible de le désigner nommément, au moins d'identifier un groupe restreint de personnes correspondant. Nos craintes peuvent paraître excessives, mais 
les conséquences, tant pour les patients que pour leurs proches, pouvant être lourdes, nous ne pouvons accepter de considérer la tâche d'anonymisation comme pleinement réalisée.

Nous travaillons donc à partir de la ressource transcrite et désidentifiée. Nous avons, dans un deuxième temps, construit une version du corpus où les tours de paroles ont été mélangés, en ne conservant que la catégorie du locuteur (psychologue - témoin - patient). Il devient alors très difficile (impossible) de reconstruire les histoires, les temporalités ou encore les géographies, et nous pouvons raisonnablement considérer que l'anonymat est garanti. 


\section{Protocole expérimental}
   \label{expe}

Nous avons annoté le corpus automatiquement en disfluences, en morpho-syntaxe et lemmes, puis nous avons réalisé une analyse textométrique outillée. Cette section présente les outils utilisés.


\subsection{Annotation automatique des disfluences : \tool{Distagger}}
  L'annotation en disfluences a été réalisée pour en étudier la pratique chez les schizophrènes (section~\ref{analysedisfluences}).

\tool{Distagger}~\cite{Constant2010} est un outil d'annotation automatique des disfluences librement disponible dont les performances ont été évaluées, sur un corpus de référence de 22~476 mots et 1~280 disfluences, à 95,5~\% de F-score (précision de 95,3~\%, rappel 95,8~\%)\footnote{Une évaluation de l'outil sur un échantillon de nos données (4 entretiens) a mis au jour un taux d'erreur compris entre 5 et 10~\%. Une analyse de ces erreurs a montré qu'elles étaient majoritairement dues à des interruptions mal identifiées, problème que nous avons corrigé depuis.}.

%
%
%
%
 De manière tout à fait classique pour la question des disfluences, \tool{Distagger} les définit comme des réalisations orales qui rompent la continuité syntaxique. Il permet d'identifier des réalisations de natures différentes, pour lesquelles quatre restent prédominantes dans les corpus oraux : les \textit{euh}, les répétitions, les autocorrections immédiates et les amorces de morphèmes. Ces différentes réalisations sont définies comme suit (les exemples proviennent de notre corpus) :

 \begin{itemize}
 \item Les \textit{euh} :

 \begin{example}
 moi ça m'est presque plus \underline{euh} difficile et \underline{euh} anti-naturel de parler
 \end{example}

 \item Les répétitions sont entendues comme la reprise explicite et identique d'un mot ou d'un groupe de mots dans le contexte immédiat d'apparition. La répétition peut contenir ou être précédée d'un mot creux comme \textit{oui}, \textit{non}, ou un \textit{euh} :
 \begin{example}
 j' arrive \underline{à être} à être concentrée quand il faut faire quelque chose
 \end{example}

 \item L'autocorrection immédiate est une variante de la répétition dans laquelle un trait morphologique peut varier (ce qui apparaît régulièrement avec les déterminants) :
 \begin{example}
 enfin je sais pas trop \underline{le} les termes
 \end{example}

 \item L'amorce est une interruption de morphème en cours d'énonciation. La fin du mot est marquée par un -.
 \begin{example}
 pis progressivement vous \underline{av-} pouvez travailler sur votre concentration
 \end{example}
 \end{itemize}

 Les annotations de \tool{Distagger} sur le corpus font apparaître sept étiquettes dont deux étiquettes spécifiques permettant de repérer les tours de parole et les interlocuteurs à qui ils sont associés. Par ailleurs, deux autres apparaissent dans des volumes trop restreints pour être significatifs (respectivement 5 et 1 étiquettes).
 Dans la suite nous utiliserons les trois étiquettes: $\{EUH\}$, $\{REP\}$ et $\{CORR\}$.
 Une remarque importante est que les amorces sont soit reconnues comme des répétitions, soit comme des corrections. 
Comme nous nous intéressons uniquement au volume de disfluences et non à leur distribution en catégories, nous conservons cette version de l'annotation. Il va de soit qu'un prolongement de notre proposition nécessiterait de revenir plus en détails sur cette distribution et sur les disfluences combinées, sans se contenter du \textit{reparandum\footnote{Le reparandum est la partie de l'énoncé précédent la disfluence et qui doit être corrigée.}}.


\subsection{Annotation automatique en morpho-syntaxe : \tool{MElt}}
\label{MElt}
\tool{MElt}~\cite{Denis2009} est un analyseur morpho-syntaxique (\textit{tagger}) librement disponible reposant sur des perceptrons multiclasses. Il est distribué avec un modèle pour le français parlé entraîné sur le corpus TCOF-POS~\cite{Benzitoun2012} et utilisant le lexique Le\textit{fff}. Les performances de cet outil avec ce modèle atteignent 97,61~\% d'exactitude, elles sont donc au niveau de l'état de l'art.

\tool{MElt} est appelé en ligne de commande, directement sur les documents textuels auxquels nous apportons des méta-données. Nous avons donc implémenté une série de scripts en \tool{Python} pour pré-traiter le corpus et lui appliquer \tool{MElt}.

Afin de conserver l'intégrité des données, nous avons fait le choix d'appliquer \tool{MElt} sur chacun des entretiens, individuellement. Du point de vue opérationnel ce choix n'apporte pas la meilleure efficacité (le temps de chargement des ressources de \tool{MElt} étant important), mais nous pouvons ainsi post-traiter chacun des entretiens.

Techniquement, nous divisons le corpus en deux fichiers, l'un contenant l'identification du locuteur, l'autre le contenu du tour de parole. Sur ce second fichier nous appelons \tool{MElt}, puis nous reconstruisons la ressource originale augmentée des annotations en fusionnant ces deux fichiers.


Les annotations ont la forme suivante, le caractère $*$ étant utilisé pour annoter les lemmes des mots inconnus du lexique~:
\begin{example}
Voilà alors peut-être vous pouvez m’e/ m’expliquer

Voilà/FNO/voilà alors/ADV/alors peut-être/ADV/peut-être vous/PRO:cls/vous pouvez/VER:pres/pouvoir m'/PRO:clo/me e/ADV/*e //MLT/*/ m'/PRO:clo/me expliquer/VER:infi/expliquer
\end{example}

\subsection{Analyse textométrique : \tool{TXM}}
\label{txm}

\tool{TXM}~\cite{heidenhalshs-00549779} est un outil d'analyse textométrique, librement disponible, de corpus textuels, incluant des fonctionnalités d'analyse statistique (\textit{via} le logiciel R).
Outre sa facilité d'utilisation, \tool{TXM} présente un avantage décisif par rapport à des logiciels d'analyse statistique générique : il offre un accès direct au contexte, ce qui permet d'affiner les résultats quantitatifs par une analyse qualitative manuelle.

Par défaut et pour des raisons de compatibilité logicielle, \tool{TXM} étiquette les corpus avec  \tool{TreeTagger}~\cite{Schmid94probabilisticpart-of-speech}, un étiqueteur morpho-syntaxique dont les performances ne sont pas au niveau état de l'art\footnote{\tool{TreeTagger} atteint 95,7~\% d'exactitude sur le français~\cite{Allauzen2008}, ce qui correspond à peu près à deux fois plus d'erreurs que \tool{MElt}.} et qui ne propose pas de modèle spécifique pour le français parlé. Nous avons donc annoté et lemmatisé le corpus avec \tool{MElt}, et l'avons importé ainsi enrichi dans \tool{TXM}, en prenant soin de neutraliser \tool{TreeTagger}.

Nous avons utilisé \tool{TXM} pour évaluer la sur-représentation ou la sous-représentation de certains mots dans les sous-corpus de schizophrènes par rapport aux sous-corpus de témoins. Nous avons pour cela créé une partition du corpus, puis nous avons calculé les spécificités~\cite{Lafon1980} de chaque lemme, ce qui permet de prendre en compte les déséquilibres entre sous-corpus.

Nous en avons profité pour réaliser un calcul de la richesse lexicale de chaque sous-groupe (schizophrènes, témoins et psychologues) qui est le ratio du nombre de lemmes par rapport au nombre total de forme et un indice de diversité lexicale, qui est le ratio du nombre de lemmes par rapport, cette fois, au nombre total de formes différentes (types).
%
%
%
%
Il est à noter que la richesse lexicale se différencie du TTR par le fait que nous prenons en compte les lemmes et non les types (qui sont des formes). La diversité lexicale permet elle de minimiser l'impact des mots très utilisés dans le calcul de la richesse lexicale. Plus la valeur est proche de 1, plus l'interlocuteur utilise de termes différents, indépendamment de leurs dérivations morphologiques.

Enfin, nous avons manuellement examiné les contextes des lemmes présentant des spécificités élevées, afin de vérifier à quoi ceux-ci correspondent dans le corpus.

\section{Résultats sur le corpus}

\subsection{Significativité}

Dans la suite de notre présentation, nous allons revenir sur plusieurs résultats calculés sur différents volumes de données. Si une lecture directe des résultats peut sembler apporter des éléments d'interprétation, nous avons choisi de valider ces interprétations en faisant appel à une mesure de significativité. Pour cela nous avons repris celle utilisée dans~\cite{Mareuil2013} pour des contextes similaires, c'est-à-dire l'analyse de disfluences dans des entretiens journalistiques.

Cette mesure permet de calculer un indice de distribution en fonction du nombre de mots entre deux catégories d'interlocuteurs. Nous le calculons pour les trois appariements possibles (psychologue/schizophrène, psychologue/témoin, témoin/schizophrène) :

\hfill $
 s = \dfrac{(p_1 - p_2)}{\sqrt{p(1-p)(\frac{1}{n_1}+\frac{1}{n_2}})}
$\hfill $\;$

où :

\begin{itemize}
\item $  p = (n_1 p_1 + n_2 p_2) / (n_1 + n_2)$
\item $n_1$ est le nombre de mots\footnote{Chaque token compte pour un mot (y compris dans les disfluences).} prononcés par la première catégorie d'interlocuteurs,
\item $n_2$ est le nombre de mots prononcés par la seconde catégorie d'interlocuteurs,
\item $p_1$ est la proportion du phénomène attribuée à la première catégorie d'interlocuteurs,
\item $p_2$ est la proportion du phénomène attribuée à la seconde catégorie d'interlocuteurs.
\end{itemize}


 Cette mesure présente l'avantage de comparer des travaux sur les disfluences du français.
 Elle cherche à déterminer si une production $p_\alpha$ est significativement différente d'une autre $p_\beta$.
 Le résultat $s$ suit une loi normale.
 L'hypothèse la plus simple est l'égalité entre les productions, hypothèse dite $H_0$.
 Nous tentons de rejeter $H_0$ si $s$ est inférieur au quantile 2,5 ou supérieur au quantile 97,5 d'une loi normale, c'est-à-dire 1,96.
%
La valeur trouvée doit donc être supérieure à 1,96 pour être considérée comme significative, avec un risque d'erreur de 5 \%.
Nous ne considérons pas les comparaisons multiples car nos résultats sont répartis sur trop peu de données.
%
%
%

\subsection{Analyse des disfluences}
\label{analysedisfluences}

Afin d'analyser les pratiques en disfluences, nous avons étudié d'une part la quantité de disfluences produites par chacun des groupes, et d'autre part la position des disfluences dans l'entretien.
\begin{center}

\begin{figure}[htp]

\includegraphics[scale=0.35]{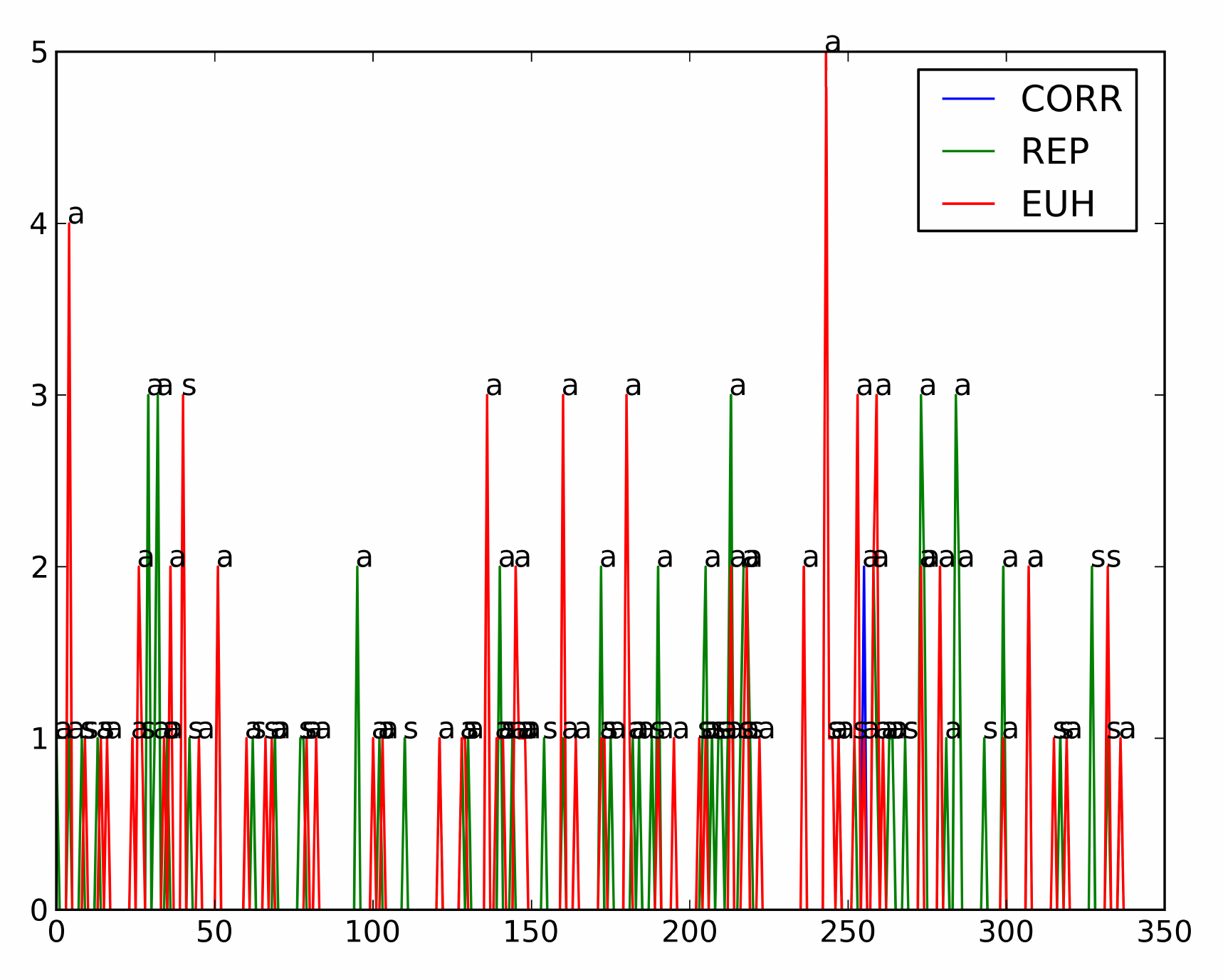} %
\includegraphics[scale=0.032]{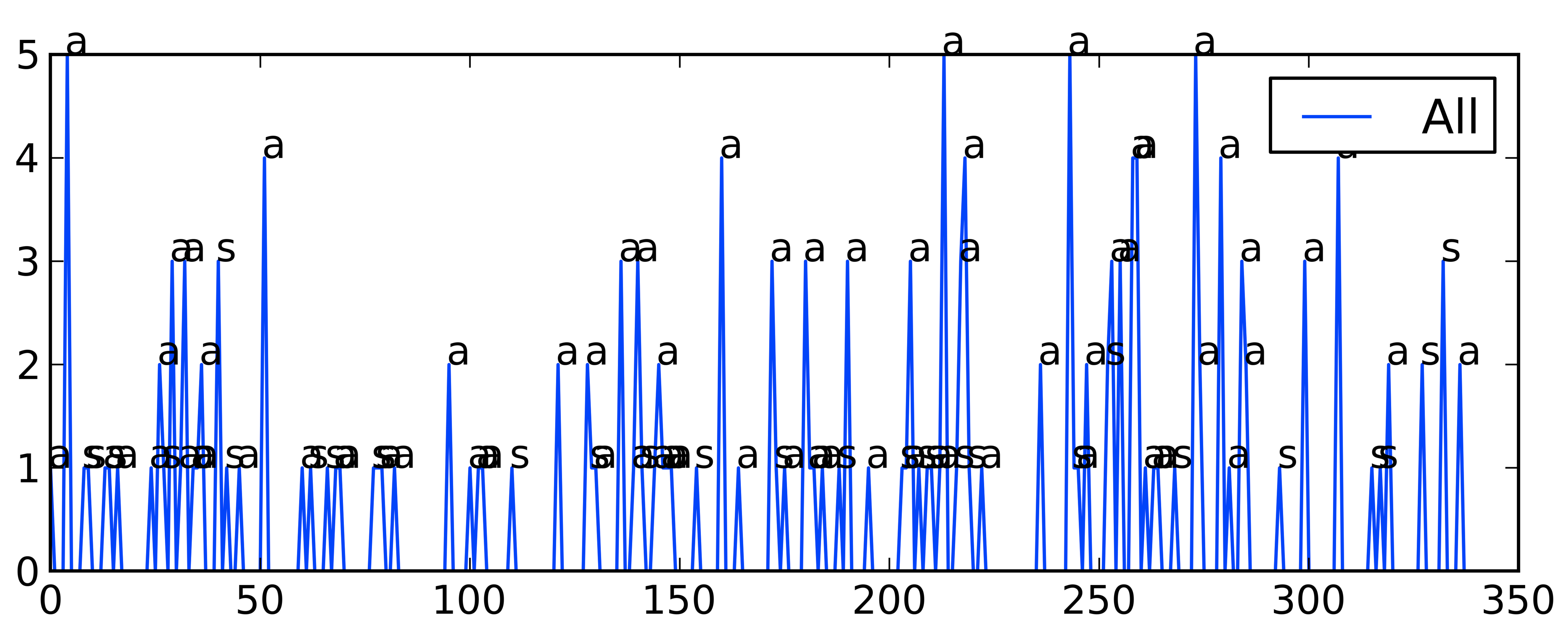}
$\quad$ \hfill (a)\hfill(b)\hfill$\;$

\caption{
Nombre d'étiquettes de disfluence par tour de parole pour un entretien du sous-corpus Ville 1. L'abscisse est la position du tour de parole dans l'entretien. Les tours de parole du psychologue sont notés par un $s$ et ceux du schizophrène par un $a$\label{fig:distributiontag}.}
\end{figure}
\vspace{-4ex}
\end{center}

La figure~\ref{fig:distributiontag} présente un exemple des résultats obtenus, pour un patient (sous-corpus Ville1). Dans la première figure ($a$), une couleur est attribuée à chacune des trois étiquettes principales, l'axe des abscisses correspond à la position du tour de parole dans l'entretien et celui des ordonnées au nombre de disfluences dans ce tour de parole. Pour les points où l'ordonnée est différente de $0$, une étiquette est apportée, $a$ pour les tours de paroles du patient, et $s$ pour le psychologue. La seconde figure présente les mêmes données, en affichant la somme du nombre d'étiquettes pour le même tour de parole. Ainsi, la première valeur significative est à 4 dans la figure $a$ et à 5 dans la figure $b$. Dans ce tour de parole \tool{Distagger} identifie 4 $EUH$ et 1 $REP$.

Ces figures, même si elles n'apportent pas directement de résultats sur les disfluences des schizophrènes, nous ont permis d'identifier un schéma récurrent sur l'ensemble des entretiens. En effet, nous pouvons visualiser deux moments de stress où les disfluences augmentent chez le schizophrène. Le premier est en début d'entretien et peut simplement s'entendre comme la tension dû à la rencontre avec le psychologue. Le second moment est plus difficile à interpréter, il intervient au bout de deux tiers de l'entretien, et ce, quelle qu'en soit la longueur. Nous pensons qu'il s'agit en fait d'un indice de fatigue que le psychologue intègre de manière implicite. Comme il ne s'agit pas de brutalement mettre fin à l'entretien, il accompagne le patient vers la fin de la rencontre dans le dernier tiers. Si on retrouve également des moments de stress dans les entretiens du groupe des témoins, cette régularité ne semble pas se confirmer. Elle appartiendrait bien au groupe des schizophrènes.

Au delà de cette première analyse, nous avons procédé à une évaluation quantitative systématique.
Nous calculons sur l'ensemble des trois étiquettes $CORR$, $REP$ et $EUH$, les fréquences d'apparition. Nous avons choisi de normaliser ces fréquences par rapport au nombre de mots prononcés en fonction de la catégorie de l'interlocuteur.
Le tableau~\ref{tab:resgene} rassemble toutes les données issues de l'analyse produite par \tool{Distagger}.
Nous présentons les résultats pour chacune des catégories (en sommant tous les résultats des psychologues, qu'ils soient devant un témoin ou un patient).
Le total reprend la somme des valeurs de la catégorie d'interlocuteur. 

\begin{table}[h]
\begin{center}
 $\begin{array}{|l|lll|lll|}
 \hline
 &\multicolumn{3}{c|}{$corpus Ville2$}&\multicolumn{3}{c|}{$corpus Ville1$}\\
 & $S$ & $T$ &  $P$  & $S$ & $T$ &  $P$ \\
\hline
CORR  & 0,0004&9e-05 &  0,0001  & 0,0013 & 0,0007 &  0,0006\\
\hline
REP  & 0,0125 & 0,0078 &  0,0067 & 0,0211 & 0,0134 &  0,0174\\
\hline
EUH  & 0,0190 & 0,0089 &  0,0073 & 0,0369 & 0,0326 &  0,0282\\
\hline
total&\textcolor{red}{0,032}&\textcolor{red}{0,0168} & \textcolor{red}{0,0142}&\textcolor{blue}{0,0595}&\textcolor{blue}{0,0468} & \textcolor{blue}{0,0463}\\
\hline
\end{array}$
\end{center}
\caption{Répartition des étiquettes de \tool{Distagger} dans les sous-corpus, normalisée par rapport au nombre de mots  (T = témoins, S = schizophrène, P = psychologue).\label{tab:resgene}}
\end{table}

La lecture du tableau montre que la production de disfluences des témoins et des psychologues sont du même ordre de grandeur : $1,68~\%$ et $1,42~\%$ pour le sous-corpus Ville2, et $4,68~\%$ et $4,63~\%$ pour le sous-corpus Ville1. Dans le même temps, les productions des schizophrènes sont supérieures : $3,2~\%$ et $5,95~\%$.
L'observation de la différence du pourcentage de disfluences entre les non-schizophrènes et les schizophrènes est alors relativement stable : $1,63~\%$ dans le sous-corpus Ville2 et $1,29~\%$ dans le sous-corpus Ville1.

La variabilité des mesures obtenues entre les deux sous-corpus peut paraître importante, mais elle nous semble venir de la qualité de la transcription. De plus, la répartition des sujets dans les deux sous-corpus est différente, ce qui peut aussi être une explication. Néanmoins, s'il n'est pas raisonnable de proposer le calcul d'un résultat pour l'ensemble du corpus, la constance de la différence de résultats conduit à notre conclusion.

Comme nous l'avons annoncé, nous avons calculé la significativité entre ces valeurs. Les résultats obtenus sont présentés dans le tableau \ref{tableSignificativite}.

\begin{table}[h]
\begin{center}

 $ \begin{array}{|l|cc|}
  \hline
 & $corpus Ville1$ & $corpus Ville2$\\
 \hline
 $T - P$ & 0,42 &   3,23 \\
$S - P$ & 10,68   & 19,42 \\
$S - T$    & 10,28   &  16,04 \\
 \hline
\end{array}$

\caption{Significativité des disfluences entre les groupes d'interlocuteurs  (T = témoins, S = schizophrène, P = psychologue).}
\label{tableSignificativite}
\end{center}
\vspace{-4ex}
\end{table}

Il apparaît que les différences entre les témoins et les psychologues sont faibles, voire non significatives, ce qui permet de rapprocher leurs comportements. Par contre, la significativité est importante (toujours supérieure à 10) dans les appariements qui comprennent des schizophrènes, ce qui nous conduit à conclure que le nombre de disfluences produites par des schizophrènes est significativement différent de celui des non-schizophrènes de l'expérimentation (psychologues et témoins).

\subsection{Analyse des catégories morpho-syntaxiques et lemmes}

L'objectif étant d'étudier la production langagière des schizophrènes, nous avons poursuivi les analyses en nous focalisant sur les catégories morpho-syntaxiques (\textit{Part-of-Speech} - POS), ainsi que sur les lemmes. Ces objets nous intéressent en ce qu'ils devraient nous apporter des indices \textit{a priori} sur la complexité de la production.
%
%
Comme nous l'avons introduit dans la section~\ref{MElt}, nous utilisons l'outil~\tool{MElt} avec un modèle adapté à l'oral pour annoter le corpus en catégories morpho-syntaxiques et en lemmes. 

Une première analyse nous a conduit à calculer la distribution moyenne des étiquettes morpho-syntaxiques dans le corpus. Le tableau~\ref{distributioncat} rassemble les données relatives au nombre moyen d'étiquettes morpho-syntaxiques et au nombre moyen de tours de paroles  dans le corpus. Nous avons calculé le ratio entre ces deux valeurs, ainsi que le nombre moyen d'étiquettes morpho-syntaxiques différentes dans chaque entretien. Nous pouvons observer que le nombre moyen de catégories par tour de parole est homogène dans le sous-corpus Ville1 contrairement au sous-corpus Ville2. Une analyse qualitative des données pour les témoins du sous-corpus Ville2 met en avant deux phénomènes : le fait que le psychologue est beaucoup moins intervenu que dans le cas des schizophrènes, et d'autre part, le fait que les échanges restent très courts par rapport à ceux réalisés avec les schizophrènes. Cela explique le chiffre de 31,78 pour les témoins, ainsi que l'augmentation observable chez les schizophrènes et la baisse chez les psychologues.
Il est important de noter que le nombre moyen d'étiquettes morpho-syntaxiques différentes utilisées par chaque type d'interlocuteur est quant à lui très stable (variation de 35 à 40, mais avec une moyenne globale cohérente dans chaque sous-corpus). Il semble que la quantité d'étiquettes morpho-syntaxiques ne soit pas ici discriminante.



\begin{table}[h]
\begin{center}
\begin{tabular}{|ll|ll|p{3ex}p{3ex}p{3ex}p{3ex}p{3ex}p{3ex}p{3ex}p{3ex}|}
\hline
&&VER & ADJ& ADV &NOM& DET& PRP& PRO& AUT &Ratio & Diff\\
\hline
& T & 10,98& 40&524& 83& 711& 297& 234& 218& 617& 785 \\
Ville1&S&13,12&38&416& 61& 537& 238& 183& 190& 497& 632 \\
&P&13,02& 39&575& 91& 795& 318& 247& 247& 634& 738\\
\hline
& T & 31,78& 35&247& 45& 173& 199& 146& 141& 265& 243\\
Ville2&S&17,32& 37& 378 &58& 243& 266& 201& 182& 440& 498\\
&P&9,42& 35& 192& 27& 135& 117& 86& 80& 231& 194\\
\hline\end{tabular}
\end{center}
\caption{
Ratio moyen du nombre de catégories par rapport au nombre de tours de parole par entretien et nombre moyen d'étiquettes différentes par entretien,
et répartition moyenne des catégories morpho-syntaxiques en grandes catégories : VERbe, ADVerbe, NOM, DÉTerminant, PRÉposition, PROnoms et AUTres\label{replat} (T = témoins, S = schizophrène, P = psychologue).\label{distributioncat}}
\vspace{-4ex}
\end{table}%

Nous nous sommes ensuite intéressés à une première forme d'étude qualitative des étiquettes morpho-syntaxiques en les classant en grandes catégories : verbe, adverbe, nom, déterminant, préposition, pronoms et autres. Le tableau~\ref{replat} présente les valeurs moyennes par type d'interlocuteurs sur ces grandes catégories.
Par exemple, la première valeur indique que les témoins ont utilisé 524 verbes en moyenne dans leur entretien dans le sous-corpus Ville1.
Il convient d'utiliser des catégories plus fines, ce que nous laissons à une étude ultérieure.

\begin{figure}[h]
\hspace{-4ex}
\begin{tabular}{ll}
\includegraphics[width = 6cm]{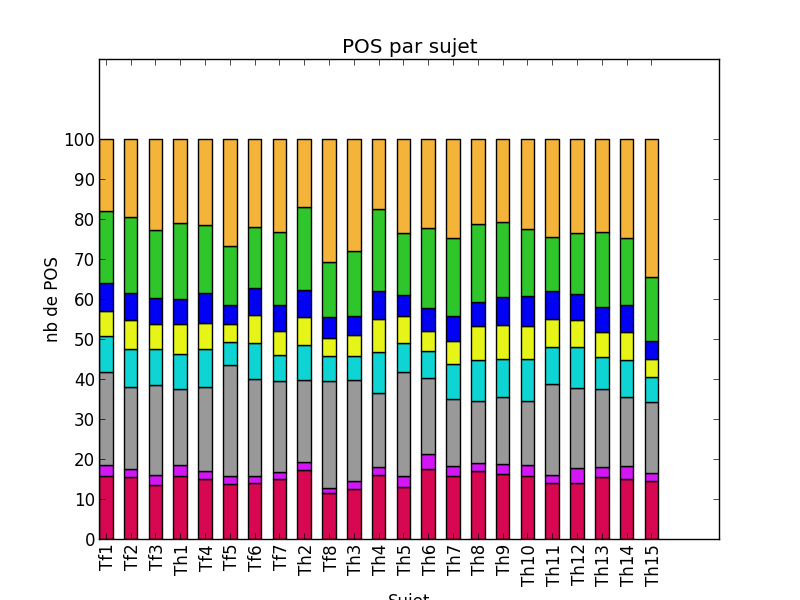} & \includegraphics[width = 6cm]{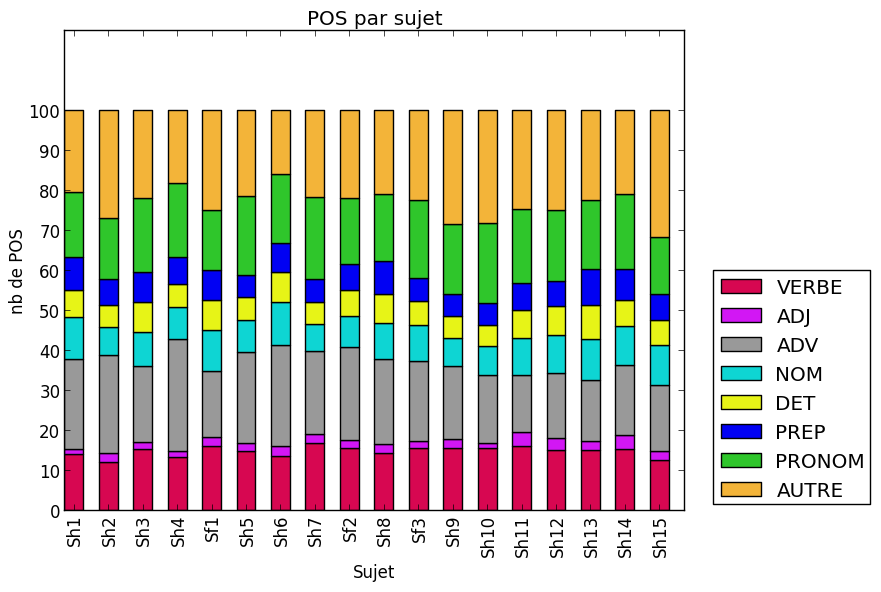} \\
\end{tabular}
\caption{Répartition des catégories morpho-syntaxiques dans le sous-corpus Ville1 pour les témoins (gauche) et les schizophrènes (droite). \label{graphiquePOS}}
\vspace{-4ex}
\end{figure}

La lecture de ces seules valeurs reste difficile, aussi avons nous produit une représentation graphique de cette répartition. 
Il apparaît que la répartition est homogène sur l'ensemble des catégories, quel que soit le type d'interlocuteur. Plus précisément, si certains entretiens contiennent une répartition divergente à celle de leur groupe, elle ne peut pas être rapprochée d'un autre comportement. La figure ~\ref{graphiquePOS} présente la répartition en grandes catégories des POS pour les témoins et les shyzophrènes du sous-corpus Ville 1.



\begin{table}[ht]

\begin{tabular}{|l|p{3ex}c|p{3ex}c|p{3ex}cp{3ex}c|p{3ex}p{3ex}p{3ex}p{3ex}|}
\hline
&  \multicolumn{2}{c|}{Ville1}& \multicolumn{10}{c|}{Ville2}\\
&&&&&\multicolumn{2}{c}{H}&\multicolumn{2}{c|}{F}&\multicolumn{2}{c}{avec trait.}&\multicolumn{2}{c|}{sans trait.}\\
&RL&DL&RL&DL&RL&DL&RL&DL&RL&DL&RL&DL\\
\hline
 T & 0,04&0.68 & 0,11&0.73 & 0,15 &0.76& 0,14&0.74&&&&\\
 S &  0,05&0.69& 0,06& 0.70&0,07&0.72&0,08& 0.71& 0,06 &0.71& 0,10&0.72\\
 P & 0,02 &0.64& 0,06&0.68&&&&&&&&\\
\hline

\hline
\end{tabular}

%

\caption{Richesse lexicale (RL) et diversité lexicale (DL) selon les sous-corpus, avec données en sexe pour le sous-corpus ville2 et en \textit{avec traitement} ou \textit{sans} pour les schizophrènes du même sous-corpus  (T = témoins, S = schizophrène, P = psychologue). 
}
\label{tab:richesseLGenerale}
\vspace{-4ex}
\end{table}

À partir de l'information extraite sur les lemmes, les formes et le nombre de mots, nous avons calculé la richesse lexicale et la diversité lexicale, comme introduites dans la section~\ref{txm}.
Le tableau \ref{tab:richesseLGenerale} montre une richesse lexicale équivalente chez les témoins et les schizophrènes dans le sous-corpus Ville1 (0,04 et 0,05, respectivement), ce qui est confirmé par la DL (0,68 et 0,69, respectivement). En revanche, dans le sous-corpus Ville2, les témoins semblent avoir une richesse lexicale supérieure (0,11 \textit{vs} 0,06 pour les schizophrènes). Or, le sous-corpus Ville2 comprend des paroles de schizophrènes qui ne sont pas sous traitement. La comparaison de leur richesse lexicale respective montre que celle des schizophrènes sans traitement est proche de celle des témoins (0,10 \textit{vs} 0,11), alors que celle des schizophrènes sous traitement est bien inférieure (0,06). Mais cela n'est pas confirmé par la DL (0,71 et 0,73).

Parmi les schizophrènes sans traitement, une très large majorité (6 sur 7) sont des femmes.
Nous avons donc voulu vérifier si la différence était liée au sexe, mais comme le montrent les résultats, ce n'est pas le cas (0,15 \textit{vs} 0,14 chez les témoins et 0,07 \textit{vs} 0,08 chez les schizophrènes).
Une explication possible des différences observées pourrait résider dans le déséquilibre important du sous-corpus Ville2, qui ne comprend que 8 témoins et 7 schizophrènes sans traitement pour 24 schizophrènes sous traitement. La mesure choisie est en effet sensible à la taille du corpus (au même titre que le TTR, voir, entre autres~\cite{Richards1987}).

La richesse lexicale des psychologues est limitée dans le sous-corpus Ville1 (RL de 0,02 et DL de 0,64, la plus faible valeur calculée pour DL) ce qui s'explique facilement par le type d'entretien réalisé. Le psychologue ne fait que maintenir l'interaction et n'emploie donc qu'un vocabulaire limité et répétitif.
Dans le sous-corpus Ville2 la richesse lexicale des psychologues est équivalente à celle des schizophrènes (RL de 0,06 et DL de 0,68) et bien inférieure à celle des témoins (RL de 0,11 et DL de 0,73), ce qui semble être cohérent avec le sous-corpus Ville1 si l'on prend en compte le déséquilibre entre témoins.

Au delà de cette analyse directe sur les données, nous avons calculé la significativité pour tous ces groupes. Le tableau~\ref{tab:richesseSignificativite} présente l'ensemble de ces résultats. Les valeurs non significatives, inférieures à 1,96, apparaissent en rouge et les valeurs moins significatives, inférieures à 10, apparaissent en bleu.

\begin{table}[h]
\begin{small}
\begin{tabular}{|l| p{3ex}p{4ex} | p{3.5ex}p{4ex} |p{3.5ex}p{3.5ex}p{3ex}p{3.5ex}|p{3ex}p{3.5ex}p{3ex}p{3.5ex}|}
\hline
&  \multicolumn{2}{c|}{Ville1}& \multicolumn{10}{c|}{Ville2}\\
&&&&&\multicolumn{2}{c}{H}&\multicolumn{2}{c|}{F}&\multicolumn{2}{c}{avec trait.}&\multicolumn{2}{c|}{sans trait.}\\

&RL&DL&RL&DL&RL&DL&RL&DL&RL&DL&RL&DL\\
\hline
T - P &21.38&18.24	&	21.14&	 11.17 	&21.70&\textcolor{blue}{7.67}	&\textcolor{blue}{8.44}&\textcolor{blue}{7.08}	&17.57 & \textcolor{blue}{8.01}	&\textcolor{blue}{4.76} & \textcolor{blue}{5.92}\\
S - P&	24.27& 20.08&	\textcolor{red}{0.14}& \textcolor{blue}{7.85} 	&\textcolor{red}{1.42}&\textcolor{blue}{6.97}	&\textcolor{blue}{4.79} &\textcolor{blue}{2.53}	&\textcolor{red}{0.61} & \textcolor{blue}{5.52}	&\textcolor{red}{0.46} & \textcolor{blue}{3.90}\\
T - S &	\textcolor{blue}{4.72}& \textcolor{blue}{4.28}&	22.70& \textcolor{blue}{6.95}	&22.19&\textcolor{blue}{4.12}	&13.79& \textcolor{blue}{5.51}	&18.56 & \textcolor{blue}{4.77}	&\textcolor{blue}{4.47}&\textcolor{blue}{2.29}\\
\hline
\end{tabular}
\end{small}

\caption{
Significativité de RL et DL pour les appariement T - S - P du tableau~\ref{tab:richesseLGenerale}  (T = témoins, S = schizophrène, P = psychologue).}
\label{tab:richesseSignificativite}
\end{table}

La première hypothèse d'une richesse et d'une diversité lexicales équivalentes entre les témoins et les schizophrènes se confirme avec des significativités de 4,72 et 4,28, à comparer avec 21,38 et 21, 27 d'une part et 18,24 et 20,08 d'autre part. Les témoins de Ville2 ont bien une richesse lexicale supérieure. La significativité entre les schizophrènes et les psychologues est très basse, 0,14, alors qu'en comparaison des témoins elle est très élevée (21,14 avec les psychologues et 22,70 avec les schizophrènes).

En ce qui concerne l'hypothèse de l'impact du traitement, la significativité poursuit sur cette interprétation. En effet, la différence avec les psychologues de ce groupe est très faible (significativité de 0,46) et constante dans la comparaison avec le groupe des témoins (4,76 entre les témoins et les psychologues et 4,47 entre les témoins et les schizophrènes sans traitement). Pour être plus précis il conviendrait de conclure que ce groupe n'a pas un comportement langagier commun (au sens où il reste significativement différent de celui du groupe témoin), mais proche du comportement attendu pour un entretien de type semi-dirigé.

Enfin, l'analyse sur le sexe est très clairement confirmée pour les hommes (très grande proximité entre les schizophrènes et les psychologues avec une significativité de 1,42, et significativité de 21,70 et 22,19 avec les psychologues et les témoins, respectivement). Les valeurs sont moins explicitent pour les femmes. Bien que les différences soient moindres, leur ordre reste le même.

L'ensemble des analyses proposées est confirmé par le calcul de la significativité.
Il apparaît que les schizophrènes n'ont pas de comportement spécifique identifié autour des catégories morpho-syntaxiques et des lemmes.

\subsection{Analyse textométrique}
\label{subsec:texto}

Nous avons souhaité approfondir l'analyse en allant au-delà de l'aspect purement quantitatif et avons pour cela utilisé l'outil de textométrie \tool{TXM}~\cite{heidenhalshs-00549779} (voir section~\ref{txm}).

\begin{figure}[ht]

\includegraphics[width=12.5cm]{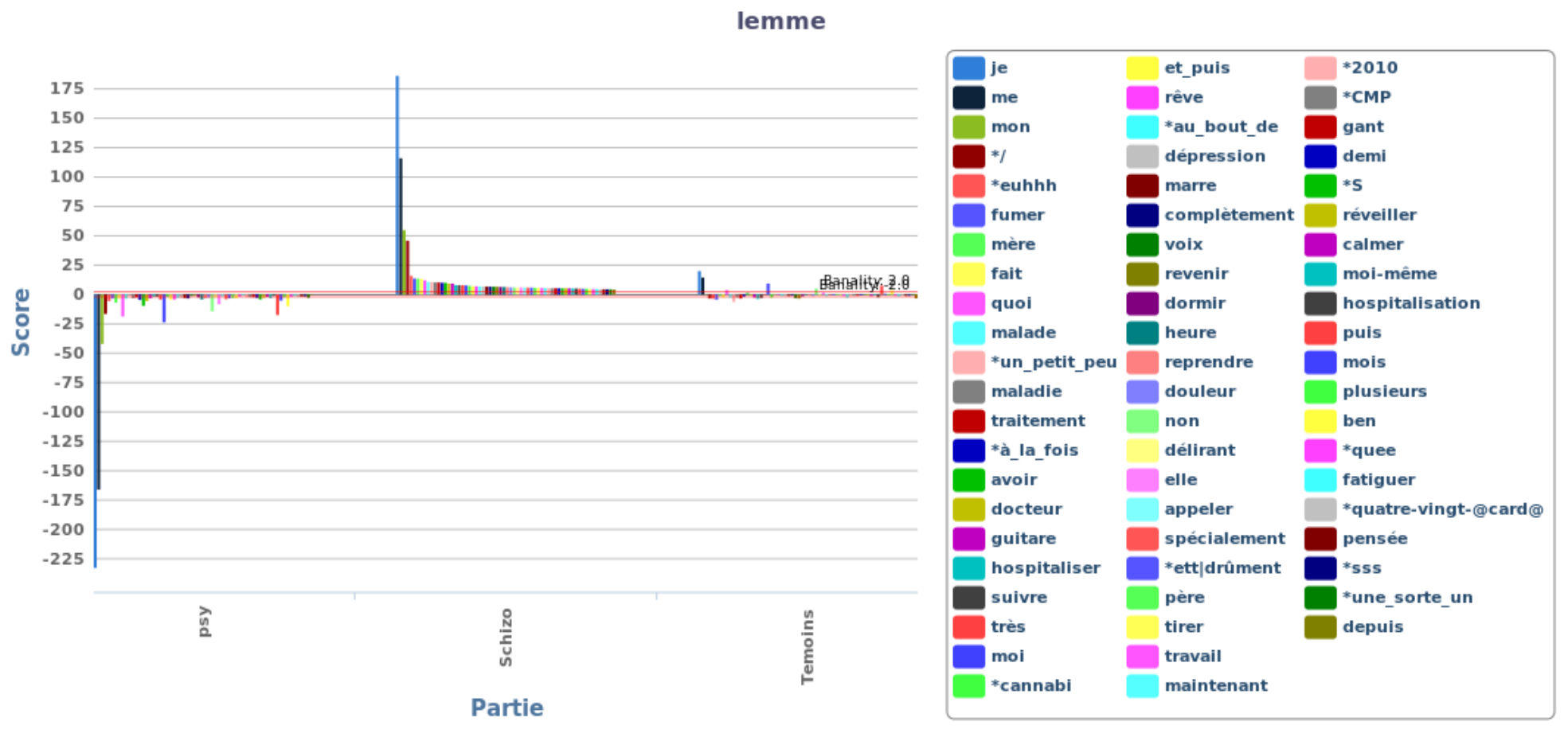}

\caption{Lemmes ayant une spécificité supérieure à 4 pour les schizophrènes du sous-corpus Ville1.\label{fig:specifVille1}}
\end{figure}

Nous avons considéré les deux sous-corpus Ville1 et Ville2 séparément, reflétant ainsi leurs différences, et avons créé pour chacun une partition selon le type de locuteur (psychologues, témoins, schizophrènes). Nous avons ensuite calculé les spécificités~\cite{Lafon1980} de chaque lemme prononcé et extrait ceux ayant une spécificité supérieure à 4, ce qui correspond à une spécificité à la fois supérieure à la zone de banalité et affichable. Les résultats de ces calculs pour les schizophrènes du sous-corpus Ville1 sont présentés dans la figure~\ref{fig:specifVille1}.

%

Au-delà des pronoms \textit{je, me, mon}, dont l'usage extensif est dû au type d'entretien, on peut voir apparaître très rapidement (en position 5 et 3) les \textit{euh} (ou \textit{euhhh}), qui participent aux disfluences\footnote{Rappelons que les lemmes commençant par une étoile correspondent à des mots inconnus du tagger \tool{MElt}.}.
Cet indice semble confirmer ce que nous avons identifié par ailleurs : les patients schizophrènes tendent à produire davantage de disfluences que les témoins.

\begin{figure}[ht]
\includegraphics[width=12cm]{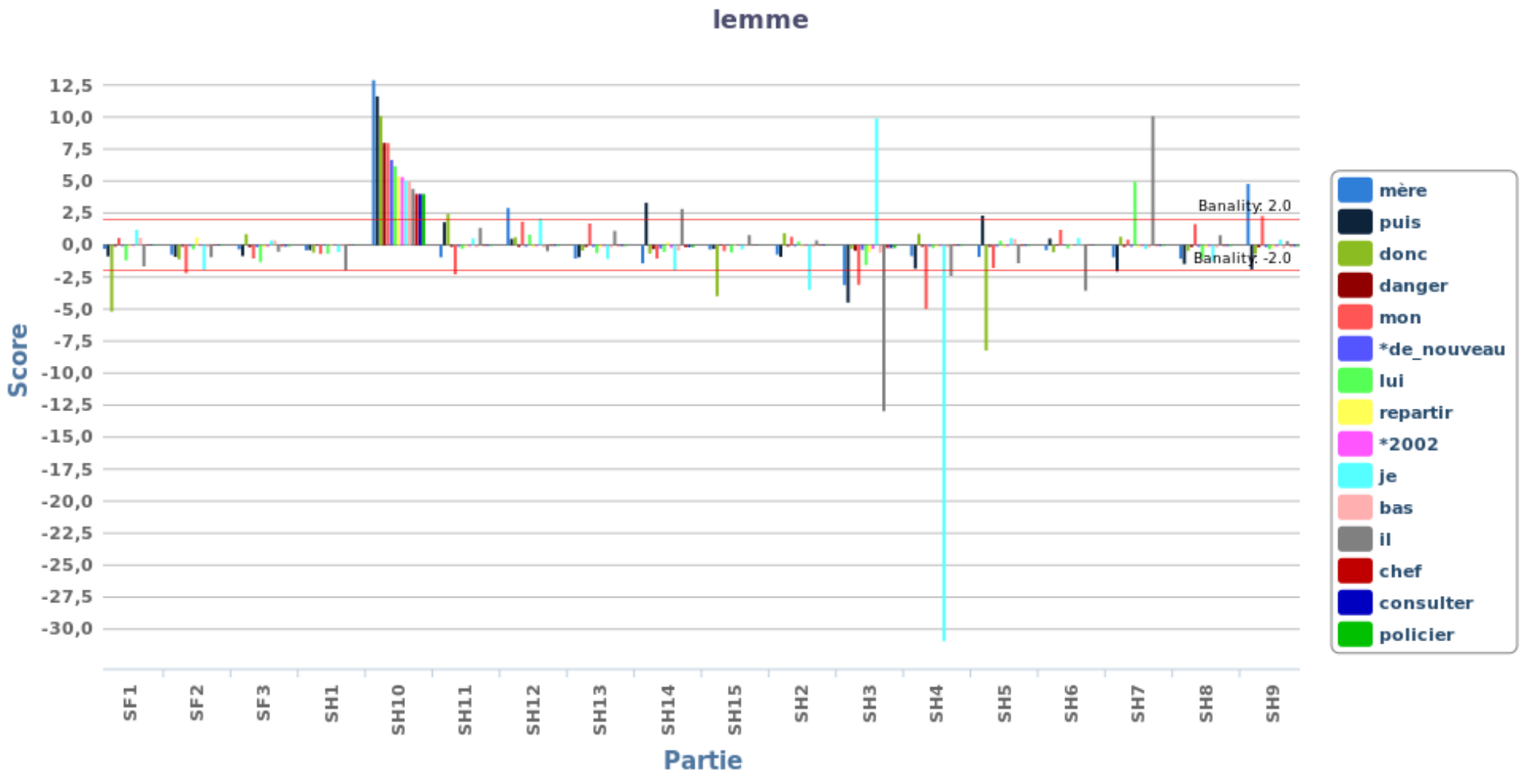}
\caption{Spécificités du lemme \textit{mère} parmi les patients schizophrènes.\label{fig:mere}}
\end{figure}

Si les données du sous-corpus Ville2 ne sont guère plus informatives, celles du sous-corpus Ville1 montrent une fréquence élevée du verbe \textit{fumer} (position 6) et, encore plus intéressant, du mot \textit{cannabis} (position 22).
Nous avons accédé aux contextes des occurrences de ces mots et avons fait plusieurs constatations : d'une part, ce n'est pas le psychologue qui incite les patients à les prononcer (ils le font spontanément), d'autre part ils sont prononcés par 6 patients différents (sur un total de 18), auxquels il faut ajouter un patient qui parle de \textit{shite}. Une explication pourrait être que, ces patients étant tous en remédiation, le personnel médical leur a probablement parlé des risques liés à la consommation de cannabis et qu'ils ont fait le lien avec leur pathologie.
En revanche, \textit{mère}, qui apparaît en position 7 dans la liste n'est pas significatif, puisqu'il est en fait employé par un seul patient (SH10), comme le montre un calcul de spécificités localisé aux patients schizophrènes (voir figure~\ref{fig:mere}).

Les termes liés à la maladie sont évidemment très présents et sont prononcés par différents patients. On trouve ainsi \textit{malade}, \textit{maladie}, \textit{traitement} (qu'il faut \textit{suivre}), \textit{douleur} (qu'il faut \textit{calmer}), \textit{docteur}, \textit{hospitaliser}, \textit{hospitalisation}. D'autres semblent plus spécifiques à la schizophrénie : \textit{dépression}, \textit{voix}, \textit{délirant}.

Cette analyse montre l'intérêt d'utiliser un outil comme \tool{TXM}, afin, non seulement d'obtenir des résultats statistiquement fiables, mais également de pouvoir accéder au contexte d'énonciation pour valider ou invalider les hypothèses émises.

   \section{Biais potentiels des expériences}
Il nous apparaît nécessaire de revenir sur différents aspects de l'étude qui participent à en biaiser les résultats. Nous pensons que ces biais ne remettent pas cause l'ensemble de la méthodologie, ni les résultats, mais ils peuvent influer sur eux. Les biais potentiels concernant l'analyse des disfluences ont été présentés de manière extensive dans~\cite{amblardhal-01054391}. Nous ne reviendrons que sur les biais communs à tout le corpus.

Une première série de biais réside dans la constitution des sous-corpus. La méthodologie ayant été largement éprouvée et améliorée pour le sous-corpus Ville1 par rapport au sous-corpus Ville2, les résultats sont plus précis sur ces données, en particulier en ce qui concerne la transcription, qui a été largement normalisée à la suite de la première collecte.

Par ailleurs, un élément important est la répartition des sujets dans l'étude. Ainsi, le sous-corpus Ville2 ne contient que 8 témoins pour 31 schizophrènes, ce qui le rend très déséquilibré. 
Nous disposons d'une autre série d'entretiens avec des témoins, réalisée à la même époque, mais dans des conditions différentes (conversations sans psychologue). Nous avons choisi d'écarter ces données qui apparaissent comme trop déviantes par rapport au reste du corpus.

\begin{table}[h]
 \centering

   $\begin{tabular}{|l|ccc|ccc|}
    \hline
& \multicolumn{3}{l|}{Schizophrènes} & \multicolumn{3}{l|}{Témoins} \\
&  & femme&homme  &  & femme&homme  \\
\hline
âge& 28,89& 30 & 28,66 & 23,22 & 22,37 &23,66\\
QI & 95,17 & 98,33 & 94,53  & 103,70 &105,5 & 102,73 \\
années d'études & 12,41& 13&12,28 & 13,17 & 13&13,26\\
  \hline
 \end{tabular}$



\caption{Moyennes des QI, du nombre d'années d'études et de l'âge des participants au corpus Ville1.}
\label{tab:qi}
\end{table}

Une analyse plus fine des sujets intégrés dans l'étude montre aussi des différences sur les capacités neuro-cognitives et l'âge des sujets. Le tableau~\ref{tab:qi} rassemble les données moyennes relatives au QI, au nombre d'années d'études et à l'âge.
On constate que les sujets schizophrènes sont significativement plus âgés que les sujets du groupe témoin avec un âge moyen de 28,89 contre 23,22. Le test de Student confirme que cette différence est significative (p=0,0058).
Par ailleurs, ils ont un QI moyen de 95,17, donc inférieur au QI moyen des témoins qui est de 103,7. La significativité issue du test de Student nous indique que cette différence fait sens (p=0,0203). Il est intéressant de noter que le nombre d'années d'études est quant à lui plutôt constant sur l'ensemble du corpus (12,41 pour les schizophrènes et 13,17 pour les témoins). 


Enfin, nous avons identifié un biais important, mais sur lequel nous ne pouvons que très difficilement influer. Les patients schizophrènes sont hospitalisés, donc sous traitement. Comme nous l'avons explicité précédemment, une très faible partie du corpus de Ville2 est constituée de patients sans traitement médicamenteux, mais, d'une part, il est très compliqué d'accéder à ces patients, et d'autre part, il n'est pas toujours évident d'interagir avec eux. Les entretiens sont généralement plus courts. Les données qui en sont issues restent pertinentes pour notre étude. La constitution d'un sous-groupe de sujets sans traitement médicamenteux, de taille équivalente au sous-groupe sous traitement n'est cependant pas réaliste.

Par ailleurs, nous n'avons pas mis en perspective la répartition de l'avancé dans la maladie des patients. Nous n'avons pas pris en compte la durée des hospitalisations, ni le fait que certains patients sont en remédiation. Pour cette dernière catégorie, cela influe directement sur leur relation à leur traitement médicamenteux (Chlorpromazine à Ville1 et neuroleptiques non spécifiés à Ville2). Les patients en remédiation et remédiation avancée sont stabilisés et prêts à être autonomisés en dehors de l'institution hospitalière. Les doses de neuroleptiques administrées peuvent être considérées comme minimales, contrairement aux patients récemment admis après une urgence qui, eux, reçoivent des doses beaucoup plus élevées.

L'influence des médicaments sur l'étude des patients schizophrènes reste une question commune à toute étude sur cette pathologie.~\cite{Levy1968} a identifié des effets négatifs (en l'occurrence, une baisse des performances) de la Chlorpromazine sur la syntaxe de quatre patients schizophrènes en calculant le ratio du nombre de propositions subordonnées produites sur la totalité des propositions produites. En outre, cet antipsychotique semble provoquer des bégaiements~\cite{Ward2008}.
Cependant,~\cite{GoldmanEisler1965} a montré (sur des sujets non schizophrènes) que les effets de cette même molécule sur les temps de pause du locuteur sont très variables selon les individus et qu'un temps de pause supérieur permet au groupe testé de générer des structures verbales complexes, comme chez les témoins. Pour ajouter à ces incertitudes,~\cite{Kremen2003} ont montré que des patients bipolaires sous antipsychotiques (dont fait partie la Chlorpromazine) présentent une meilleure fluence sémantique que les témoins.
Une réserve sur ce contre-argument doit être apportée sur la capacité des médicaments qui ne cessent de faire des progrès. Les effets secondaires ont été considérablement réduits dans les 50 dernières années. Il reste particulièrement délicat de séparer ce qui appartient à l'influence du traitement médicamenteux de ce qui est propre à celle de la pathologie.


\section{Conclusions et perspectives}


Cet article s'inscrit dans un projet plus large sur l'étude des pratiques langagières des schizophrènes. Il a été montré que ces derniers présentaient un dysfonctionnement dans la gestion de la planification du discours~\cite{MusiolTrognon96,verhaegen2007}.
L'interprétation de ce trouble se situe au niveau sémantico-pragmatique~\cite{rebuschihal-00910725} et~\cite{musiolhal-00910701}, ce qui a conduit à proposer des modélisation inspirée de la SDRT (\textit{Segmented Discourse Representation Theory})~\cite{asher2003logics}.
Cette manifestation nous renseigne sur le fonctionnement cognitif, en particulier dans son rapport à l'expression de la pensée par le langage.
Il nous est apparu nécessaire d'interroger d'autres niveaux d'analyse linguistique. Une étude manuelle n'étant pas réaliste, nous avons choisi de travailler à partir des résultats fournis par des outils de TAL existants.

Nous avons, dans un premier temps, mis en avant un usage spécifique des disfluences chez les sujets schizophrènes grâce à l'outil \tool{Distagger}. En effet, les schizophrènes produisent, respectivement dans chaque corpus,  1,63~\% et 1,29~\% plus de disfluences (par rapport au nombre de mots) que des sujets témoins ou les psychologues. Ce résultat est confirmé par un calcul de significativité.

Dans un deuxième temps, nous nous sommes intéressés aux productions en catégories morpho-syntaxiques et en lemmes. Il apparaît en effet régulièrement dans la littérature que les patients schizophrènes auraient une capacité morpho-syntaxique et une diversité lexicale réduites par rapport aux sujets témoins. Cependant, notre étude tend à montrer le contraire. Les sujets schizophrènes produisent autant de catégories morpho-syntaxiques que les autres, avec une diversité similaire. Par ailleurs, leur richesse lexicale est également similaire. Plus exactement, à partir de nos données, nous ne pouvons pas identifier de sous-classe particulière, certains sujets ayant des comportements singuliers pour chaque type d'interlocuteurs. Ces indices nous permettent de conclure que ces niveaux n'interviennent pas dans la défaillance cognitive que nous cherchons à circonscrire. Il ne s'agit alors pas ici d'une défaillance de la capacité, mais bien une défaillance de la gestion de l'expression. Le résultat précédent sur les disfluences s'inscrit tout à fait dans cette perspective.

Nous nous inscrivons donc en faux contre les précédents résultats. Il faut noter que notre corpus est significativement plus important que tous ceux utilisés dans les études auxquelles il est fait mention, qui ne dépassent jamais plus de 20 patients, alors que nous en étudions 49.

Nous avons poursuivi par une étude plus qualitative de la diversité lexicale avec l'outil~\tool{TXM}. Nous avons pu mettre en avant certaines thématiques particulières qui peuvent s'interpréter relativement facilement en fonction du contexte des entretiens. En effet, les patients schizophrènes sont en milieu hospitalier et ont une habitude des entretiens semi-dirigés. Ils ont tendance à aborder des thématiques plus classiques pour la psychologie et la psychiatrie comme leur rapport aux médicaments, à leurs dépendances, ou à leur famille.


Les prolongements de nos travaux sont de deux ordres. D'un côté nous souhaitons revenir sur le niveau de granularité de notre étude et de l'autre nous souhaitons inscrire ces résultats dans une perspective plus globale. En effet, pour les disfluences, il conviendrait d'étudier leurs lieux d'apparition à l'intérieur des tours de parole, leur dynamique dans l'interaction et de proposer des catégories plus précises (disfluences combinées, notamment). De manière similaire, pour les POS et les lemmes, il serait intéressant de regarder des catégories plus fines, en particulier concernant les verbes.

Un autre volet de l'analyse s'intéressera plus particulièrement à corréler les défaillances identifiées à la manifestation d'indices cognitifs. Bien évidemment il conviendra de les associer aux différents tests neuro-cognitifs, mais plus certainement aux comportements enregistrés par le double système oculométrique (\textit{Eye-tracking}), ainsi qu'aux enregistrements de l'activité de l'encéphale (EEG).

Les outils de TAL sont pour nous les seuls à pouvoir proposer une cartographie précise de la manifestation de ces troubles. Ils nous permettent d'une part de produire une ressource normalisée riche en méta-données (dont le manque et l'importance sont mis en valeur dans~\cite{Ghio2006}). Cependant, il apparaît très complexe de parvenir à constituer et gérer une telle ressource qui pose de nombreux problèmes éthiques.

Nous souhaiterions mettre ce corpus à disposition des chercheur(e)s, au moins une version  manuellement anonymisée et randomisée par tour de parole, mais la non planification de cette étape dans les protocoles validés par la CPP  rend cela impossible.

  \acknowledgements{
  Nous tenons à remercier les relecteurs de la revue, qui, grâce à leurs remarques constructives et détaillées, nous ont permis d'améliorer de manière significative (non calculée ici) la clarté de cet article.}

\bibliography{biblio}

\end{document}